\documentclass[a4paper,twocolumn]{article}
\usepackage{IWAC2024_template_latex}
\usepackage{amsmath}
\usepackage{amssymb}
\usepackage{amsfonts}
\usepackage{overcite}
\usepackage{arydshln}
\usepackage{bm}
\usepackage{newtxtext,newtxmath}
\usepackage[affil-it]{authblk}
\usepackage[setpagesize=false]{hyperref}
\usepackage{booktabs}
\usepackage{fancyhdr}
\usepackage{enumitem}
\usepackage{caption}
\usepackage[top=24.2truemm,bottom=21.8truemm,left=14.1truemm,right=19.9truemm]{geometry}
\usepackage{tikz}
\usetikzlibrary{positioning}

\usepackage{graphicx}

\setlist[itemize]{topsep=-1ex, parsep=-1ex}
\setlist[enumerate]{topsep=-1ex, parsep=-1ex}
\makeatletter
\def\fnum@figure{Fig. \thefigure. }
\def\fnum@table{Table \thetable. }
\makeatother

\makeatletter 
\def\@citess#1{\textsuperscript{#1)}} 
\makeatother
\usepackage{setspace}

\makeatletter
\renewcommand{\@biblabel}[1]{#1)}   
\makeatother

\makeatletter
\def\section{\@startsection{section}{1}{\z@}{0.8ex plus 1.0ex minus 0ex}{0.8ex plus 1ex minus 0ex}{\normalsize\bf}}
\def\subsection{\@startsection{subsection}{2}{\z@}{0.0ex plus -0.1ex minus 0.0ex}{0.1ex plus -0.5ex}{\normalsize\bf}}
\def\subsubsection{\@startsection{subsubsection}{3}{\z@}{0.5ex plus 0.5ex minus 0.2ex}{0.5ex plus 0.2ex}{\normalsize\bf}}
\makeatother

\makeatletter
\def\@listi{%
  \setlength{\leftmargin}{\leftmargini}%
  \setlength{\parsep}{0pt}%
  \setlength{\topsep}{0.5\baselineskip}%
  \setlength{\itemsep}{0pt}%
}
\makeatother

\renewcommand{\headrulewidth}{0pt}

\title{\fontsize{14pt}{14pt}\selectfont Effective Management of Airport Security Queues with Passenger Reassignment}

\author[1)$\dagger$]{\normalsize Shangqing Cao}
\author[1)]{\normalsize Aparimit Kasliwal}
\author[1) 2)]{\normalsize Masoud Reihanifar}
\author[2)]{\normalsize Francesc Robusté}
\author[1)]{\normalsize Mark Hansen}
\affil[1)]{\footnotesize Department of Civil and Environmental Engineering, University of California, Berkeley, USA}
\affil[2)]{\footnotesize Barcelona Innovative Transportation (BIT), Technical University of Catalonia BarcelonaTech (UPC), Barcelona, Spain}
\affil[$\dagger$]{\footnotesize email: caoalbert@berkeley.edu}

\abstract{
Airport security queues often suffer from inefficiencies that result in long wait times and decreased throughput, especially at peak departure time, affecting both passengers and airlines. This work addresses the problem of reassigning passengers to specific time slots for crossing security, aiming to mitigate these inefficiencies. We frame this problem as a Minimum Cost Network Flow (MCNF) problem, enabling us to solve it exactly in polynomial time due to its linear programming structure. Our approach redistributes passenger demand across different time intervals. By optimizing the reassignment of passengers to $\delta$-minute time slots, we achieve significant improvements in throughput and reductions in waiting time. Preliminary results demonstrate the effectiveness of our method in enhancing operational efficiency and passenger satisfaction. The MCNF formulation offers a scalable and adaptable solution, providing long-term benefits for airport security management.
}


\begin{document}

\maketitle
\thispagestyle{fancy}
\normalsize

\setlength{\abovedisplayskip}{-2ex}
\setlength{\belowdisplayskip}{-0.5ex}
\setlength{\abovedisplayshortskip}{-2ex}
\setlength{\belowdisplayshortskip}{-0.5ex}

\section{Introduction}

Airports worldwide face significant challenges in managing security queues efficiently, leading to prolonged wait times for passengers and decreased throughput for airlines. These inefficiencies create a ripple effect, causing time-related costs and frustration for both passengers and airlines. Passengers experience delays, missed flights, and increased stress, while airlines face scheduling disruptions and potential financial losses \cite{bbc}. The costs associated with these inefficiencies are multifaceted. From the passengers' perspective, extended wait times translate to lost productivity, missed flights, and heightened travel anxiety. For airlines, the costs manifest as increased operational expenses, potential loss of customer loyalty, and decreased revenue from delayed or missed flights \cite{Rosenow}. 

Just before the COVID-19 pandemic, long airport security queues caused one in every seven travelers to miss a flight in the past 12 months \cite{forbes}. Nearly half of respondents in an AirHelp survey lost money due to non-refundable costs caused by airport disruptions, including long delays and missed flights \cite{fodors}. In the meantime, the airline industry aims to set a new passenger record in 2024, connecting nearly five billion people over 22,000 routes with 39 million flights \cite{iata}. Thus, addressing the long waiting time in queues at airport security control is crucial to ensuring smooth operations at airports worldwide in the face of the continuing growth in passenger demand.

One issue associated with airport security management is that passengers have varying preference in the time of arrival at airports prior to a flight. Passengers who arrive hours prior to the departure of their flights place additional stress on the security system in serving those who need more immediate security checks \cite{independent}. Studies have examined the benefits of security lane reassignment but such mitigation does not address the temporal variation in passengers' arrivals \cite{marshall}. Therefore, in this paper, we seek to address the issue of long wait time at security control by reassigning passengers to specific time slots to reach system optimum.

\section{Literature Review}

Interestingly, this problem shares similarities with the morning commute problem, where commuters are incentivized to travel early to avoid peak congestion periods. \cite{small} \cite{vickrey} In both scenarios, the goal is to redistribute demand to flatten congestion peaks, thereby enhancing overall system efficiency. For instance, just as commuters are encouraged to start their journeys earlier to reduce travel time in some cases, passengers can be incentivized to pass through airport security during less busy periods. However, in the morning commute problem, the cost can be broken down into two types: travel time penalty and congestion cost \cite{lamotte}. It is assumed that at equilibrium, the first and the last user going through the same bottleneck would incur the same cost. However, in the airport security setting, there exists an abrupt increase in cost when a user's time of arrival is past the flight departure time. It is difficult to reach the system equilibrium because passengers in the air transportation system do not usually have the opportunity to repeatedly assess the cost associated with different time of arrival prior to a flight as most people travel by air only a few times a year.

This problem can also be likened to the Single-Sourced Capacitated Facility Location Problem (SSCFLP) \cite{odoni1} or a Capacitated Facility Location Problem (CFLP) both of which involve allocating limited resources (in this case, security slots) to meet varying demand levels. The idea of optimizing the reassignment of demand to proper supply providers is used extensively in the design of control policies for operating air transportation systems. \cite{ang} \cite{leizhou}.

Our approach models the problem as a Minimum Cost Network Flow (MCNF) problem, which allows for an exact polynomial-time solution. This distinction highlights the flexibility and efficiency of the MCNF approach in addressing the dynamic and time-sensitive nature of airport security queue management. We minimize the total cost associated with security queue congestion, balancing demand across different time intervals. Preliminary results from our study indicate significant improvements in throughput and reduced wait times for passengers. Our approach not only addresses the immediate inefficiencies in security line management but also offers a scalable solution that can adapt to varying demand patterns, providing long-term benefits for airports, airlines, and passengers alike.

The rest of the paper is organized as follows. Section \ref{section:method} details the mathematical formulation of the MCNF problem within the context of security reassignment. In section \ref{section:sqs}, we establish a queuing system for the airport security screening process and define the metrics on which we are evaluating our control policy. Section \ref{section:results} showcases the results under both a deterministic, nominal condition, and under variation in passenger compliance with the assignment.

\section{Methodology}
\label{section:method}

We consider the problem of reassigning passengers to $\delta$-minute time slots and $\delta$ denotes the interval length of each time slot. These time slots represent the assigned time at which passengers cross security at the airport. We use the phrase "reassigning" to differentiate from the current first-come first-served basis. We frame this problem as a Minimum Cost Network Flow (MCNF) problem, which is well-studied in the optimization and operations research literature. While this problem can be formulated in multiple ways, including as a Single-Sourced Capacitated Facility Location Problem (SSCFLP), the MCNF formulation can be solved exactly in polynomial time using network Simplex with integral solutions.

\tikzset{%
  every neuron/.style={
    circle,
    draw,
    minimum size=1cm
  },
  neuron missing/.style={
    draw=none, 
    scale=4,
    text height=0.333cm,
    execute at begin node=\color{black}$\vdots$
  },
}
\begin{figure}[h!]
\begin{tikzpicture}[x=1.5cm, y=1.5cm, >=stealth]

\foreach \m [count=\y] in {1,2,3,missing,4}
  \node [every neuron/.try, neuron \m/.try] (input-\m) at (0,2.5-\y) {};

\foreach \m [count=\y] in {1,2,3,missing,4}
  \node [every neuron/.try, neuron \m/.try ] (hidden-\m) at (2,2.5-\y) {};

\foreach \m [count=\y] in {1}
  \node [every neuron/.try, neuron \m/.try ] (output-\m) at (4,-0.5) {};

\foreach \l [count=\i] in {P1,P2,P3,PP}
  \node [left] at (input-\i.west) {\l};

\foreach \l [count=\i] in {T1,T2,T3,TT}
  \node [left] at (hidden-\i.west) {\l};

\node [left] at (output-1.west) {Sink Node};

\foreach \i in {1,...,4}
  \foreach \j in {1,...,4}
    \draw [->] (input-\i) -- (hidden-\j);

\foreach \i in {1,...,4}
  \draw [->] (hidden-\i) -- (output-1);

\foreach \l [count=\x from 0] in {Passengers, Time Slots, Sink Node}
  \node [align=center, above] at (\x*1.75,2) {\l \\};

\end{tikzpicture}

\caption{Flow network representation of passengers and time slots with a sink node. Arrows denote feasible flows from passengers to time slots and from time slots to the sink node.}
\label{flownetwork}
\end{figure}
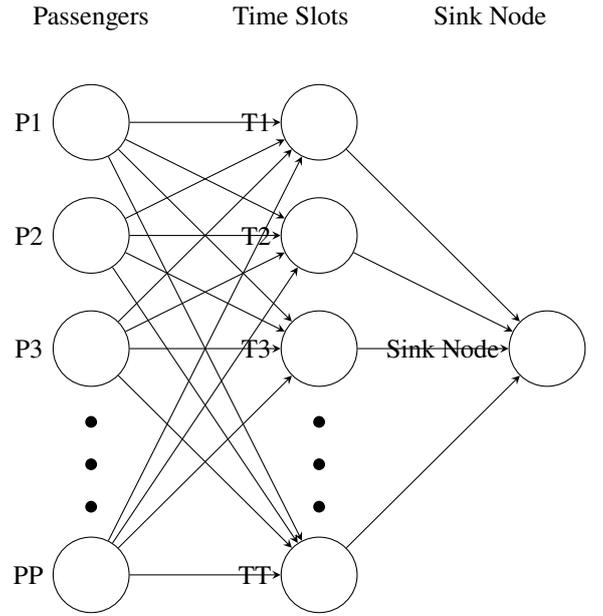

\subsection{MCNF Formulation}

We construct a directed graph $G(V,E)$ where $V$ is the set of nodes and $E$ is the set of edges in the network. Our goal is to find the cheapest assignment of the flow in this network $G$ while satisfying the flow balance constraints and the capacity constraints. We define the network as the following:

Let $p \in P$ be a passenger to be reassigned, where $P$ is the set of passenger nodes. Let $t \in T$ be a time slot to which the passenger can be reassigned, where $T$ is the set of time slots. Let $C_{t}$ denote the capacity of the system in time slot $t$. We define $V = P \cup T \cup \{s\}$, where $s$ denotes the sink node and we define $E = \{e_{pt} | p \in P, t \in T\} \cup \{e_{ts} | t \in T\}$. Figure \ref{flownetwork} visualizes the network we construct.

The decision variables in our formulation are the $x_{ij}$s, which are the flow on the directed edges connecting node $i$ to node $j$, and $(i,j) \in E$. The respective unit cost of flow on each edge is $c_{ij}$. We use $l_{ij}$ and $u_{ij}$ to denote the lower and upper bounds of flow on edge $(i,j) \in E$. We use $b_{v}$ to denote the net flow of each node $v\in V$.

We adopt a standard MCNF formulation where we seek to minimize the weighted sum of flow on all edges while ensuring that the net flow requirement with respect to each node is satisfied (Eq.\ref{const1}), and that the flow on each edge is within the range we define (Eq.\ref{const2}). 

\vspace{0.1in}

\begin{align}
\text{min} \sum_{i}\sum_{j} c_{ij}x_{ij}
\end{align}
\vspace{0.1in}

\begin{align}
    \sum_{k\in V \setminus \{v\}} x_{ik} - \sum_{k\in V \setminus \{v\}} x_{ki} = b_{i} \quad \forall v \in V
    \label{const1}
\end{align}
\vspace{0.3in}
\begin{align}
    l_{ij} \leq x_{ij} \leq u_{ij} \quad \forall (i,j) \in E
    \label{const2}
\end{align}

\vspace{0.1in}
We now define the parameters $b_{i}$, $l_{ij}$, and $u_{ij}$.
\vspace{0.1in}

\begin{align}
    b_{i} = \begin{cases}
        1 & \text{if } i \in P \\
        0 & \text{if } i \in T \\
        -|P| & \text{if } i = s
    \end{cases}
    \label{bi}
\end{align}

\vspace{0.3in}

\begin{align}
    l_{ij} = 0 \quad \forall (ij) \in E
    \label{lij}
\end{align}
\vspace{0.1in}

\begin{align}
    u_{ij} = \begin{cases}
        1 & \text{if } i \in P, j \in T \\
        C_{i} & \text{if } i \in T, j = s
    \end{cases}
    \label{uij}
\end{align}
\vspace{0.1in}

Eq. (\ref{bi}) ensures that each passenger is served and Eq. (\ref{uij}) ensures that the number of passengers we reassign to each time slot is within the capacity of the particular time slot. 

\subsection{Cost Function}
\label{cost_function}

We assume that passengers arrive at the security queue an hour before their flight departure time. We establish this assumption so that we can compare the solution output from the optimization problem to the queuing system under the first-come first-served basis outlined in \ref{section:sqs}. The arrival process of passengers can be substituted by different assumptions and that the computational complexity of the model does not depend on the assumption we establish. 

Assigning a passenger to a time slot one hour prior to the flight departure time incurs a cost of 0. We assume a quadratic cost structure for reassigning passengers to time slots more than 1 hour before the flight departure time, and a linear cost structure for reassigning passengers to time slot less than 1 hour before the flight departure time but before the flight departure time. The difference in the order of the cost function assumes that it is more feasible to delay a passenger's entrance to the security queue compared to asking the passenger to arrive early. If the decision is to assign a passenger to a time slot after the flight departs, we then set the cost to a sufficiently large number to penalize missing the flight. Let $t_{i}$ be the flight departure time of passenger $i$ and then the cost function is defined as a piece-wise function:

\begin{align}
    c_{ij} = \begin{cases}
        0 & \text{if } j = t_{i} \\
        \alpha \cdot (j - t_{i}) & \text{if } j \in (t_{i}, t_{i}+1] \\
        \beta \cdot (t_{i} - j)^{2} & \text{if } j < t_{i} \\
        \gamma & \text{if } j > t_{i} + 1
    \end{cases}
\end{align}

\begin{figure}[h]
    \centering
    \includegraphics[width=8cm, height=8cm, keepaspectratio, clip]{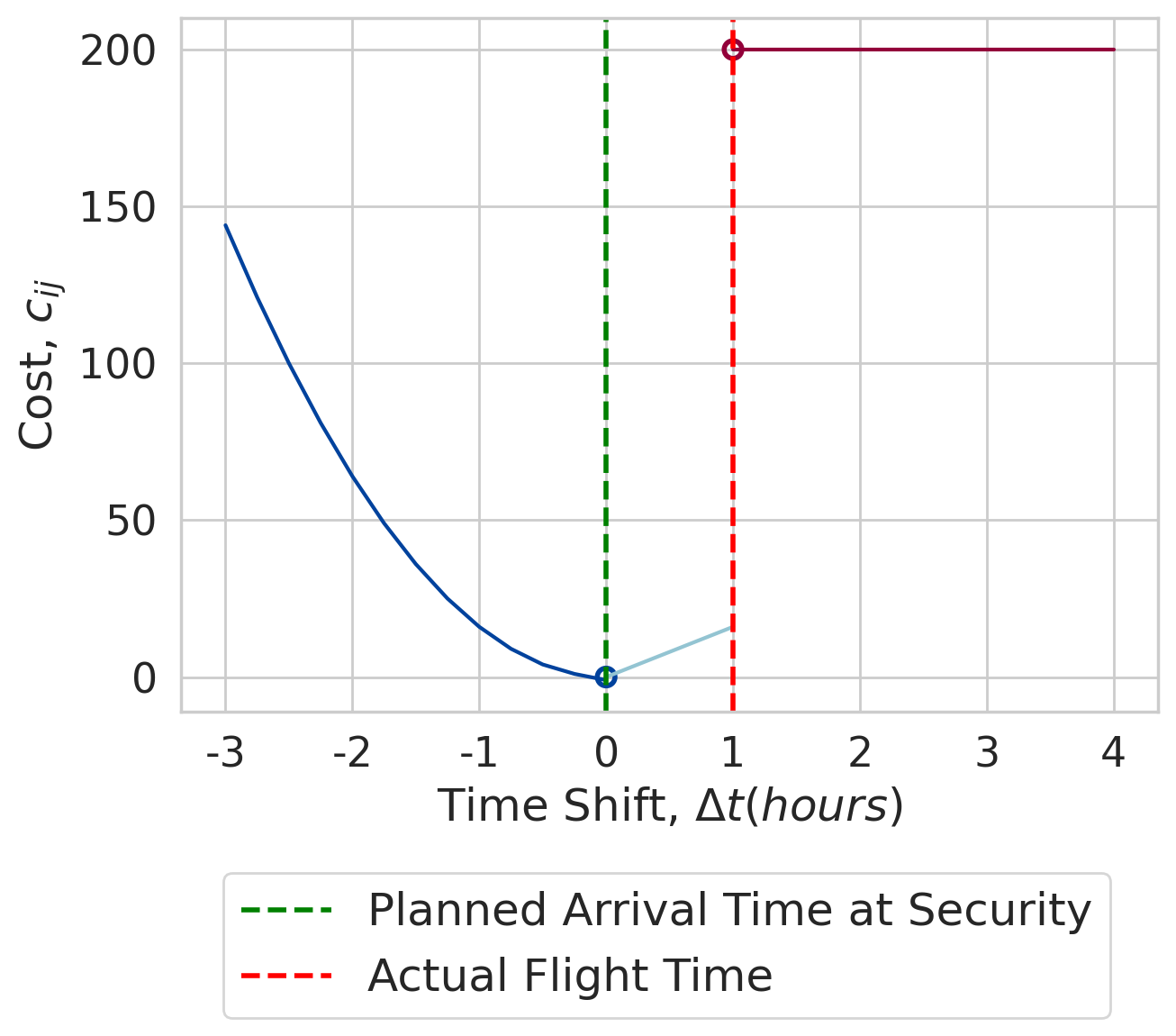}
    \caption{The cost of reassigning passengers to different time 
    slots with respect to the passenger's flight departure time.}
    \label{fig:cost}
\end{figure}

\vspace{0.1in}

where $\alpha$, $\beta$, and $\gamma$ are the coefficients. Figure \ref{fig:cost} visualizes the cost structure with respect to each individual passenger.

\section{Security Queuing System}
\label{section:sqs}

The processing times at the security queue at the airport are quintessentially determined by the staffing capacity. Understandably, the key trade-off that we are trying to address by solving this problem is to achieve the maximal gain in passenger experience, by reducing the processing times at the security queue, while ensuring a reasonable distribution of the capacities of the security queues. Note that given a certain time at which a passenger arrives at the airport security queue, the time at which they depart from the security queue is directly dependent on the capacity of the security system at that point of time. 

We establish a first-come first-served queuing system as the baseline to which our model is compared. We can view the queuing system as a passive control policy, where the act of waiting in queue is the assignment. We introduce the following notations:

\begin{itemize}
    \item $v(t)$: cumulative number of passengers who have arrived at the security queue by time $t$.
    \item $d(t)$: cumulative number of passengers who have departed from the security by time $t$.
    \item $q(t)$: cumulative number of passengers who have departed from the airport by time $t$. 
\end{itemize}  

We then use $\hat{v}(t)$, and $\hat{d}(t)$, to represent queuing system under the control policy output from the optimization problem. However, $q(t)$ remains unchanged as it is dependent on the flight schedules. Note that the function $v(t)$ is essentially the control policy represented by the decision variables in the optimization problem and $d(t)$ is a function dependent on $v(t)$ and the assumed capacity of the security system.

We use two metrics to evaluate the efficiency of the control policy derived from the optimization problem. The first metric is the total waiting $(TW)$ time in the system, as defined by:

\begin{align}
    TW = \int_{0}^{T} v(t) - d(t) dt \approx \sum_{t=1}^{T} (v(t) - d(t))
\end{align}

\vspace{0.1in}

The second metric is the total cost ($TC$) of the system as defined by the cost function explained in section \ref{cost_function}. For a system with no optimized passenger reassignment, $TC = \alpha \cdot TW$, where $\alpha$ is the time coefficient of assigning passengers to a time slot within the 1 hour period before flight departure, as shown in fig \ref{fig:cost}. The $TC$ of the optimized control is simply the optimized objective function value of the optimization problem. $TC$ captures the relative difference in unit cost and the added benefits of asking passengers to arrive early while taking the quadratic nature of the cost of assigning passengers to an earlier time slot into account.

\section{Numeric Study}
\label{section:results}

We obtained the May 9th, 2023 departure flight schedules of terminal 1 at Josep Tarradellas Barcelona-El Prat Airport (BCN) in the city of Barcelona, Spain. In our data sample, there are 260 departing flights out of BCN, with 49,034 seats. We set the length of the time interval $\delta$ to 15 minutes, which results in partitioning the day into 96 time slots. For simplicity, we assume a load factor of 100$\%$, which gives a total demand of 49,034 passengers for security processing. We assume $\alpha = 4$, $\beta = 1$, and $\gamma = 200$ for the cost function.

\subsection{Time-Invariant Capacity}
\label{sec:const_cap}
\vspace{0.1in}

For the simplest case, we assume that the capacity of the security system stays constant throughout the day. As Eq. (\ref{const1}) ensures that each passenger receives an assignment, the optimization problem becomes infeasible if we set the constant value of the capacity below a certain critical threshold. The critical capacity, $c_{c}$, is thus:

\vspace{0.1in}

\begin{align}
    c_{c} = \frac{N}{T},
\end{align}

\vspace{0.1in}

where $N$ is the total number of passengers and $T$ is the total number of time slots. Plugging in the values of the total number of passengers ($N = 49,034$) and the total number of time slots ($T = 96$), we get the critical value of the capacity to be 511.

\begin{figure}[h]
    \centering
\includegraphics[width=8cm, height=8cm, keepaspectratio, clip]{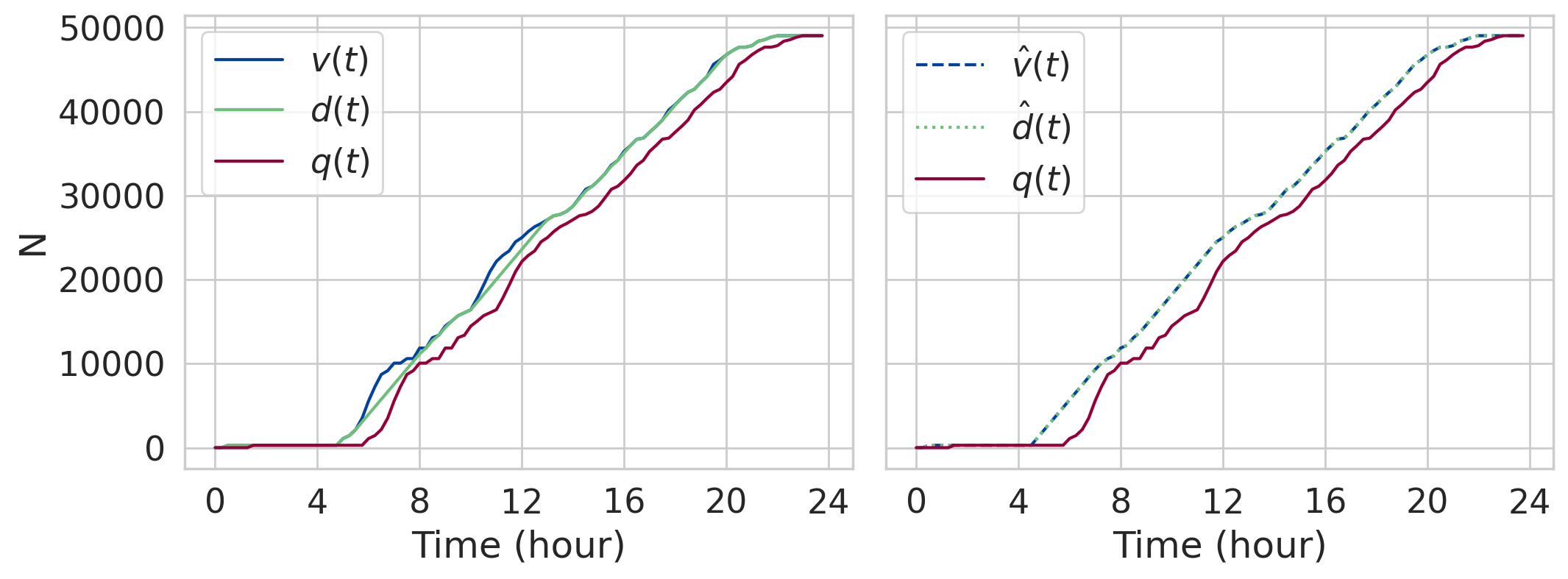}
    \caption{N-T diagram of the system under first-come first-served control (left) and under the optimal control policy (right) derived from the optimization where the capacity is not allowed to vary with time.}
    \label{fig:nominal}
\end{figure}

For a nominal case, we assume a security capacity of 900 passengers per 15-minute time slot. Figure \ref{fig:nominal} shows the queuing system before and after the reassignment of passengers. We can see that with no reassignment, a queue builds up at around 6 am and 9 am due to the early morning peak in the number of departing flights. Passengers spend up to 1 hour waiting to go through security which results in many of them barely being on time for departures. With the reassignment, which is shown on the right side in Fig. \ref{fig:nominal}, the waiting time of passengers goes to 0. The respective $TC$ for the two systems are 132,628 and 39,622, showing a reduction of roughly 70\%.

\begin{figure}[h]
    \centering
    \includegraphics[width=8cm, height=8cm, keepaspectratio, clip]{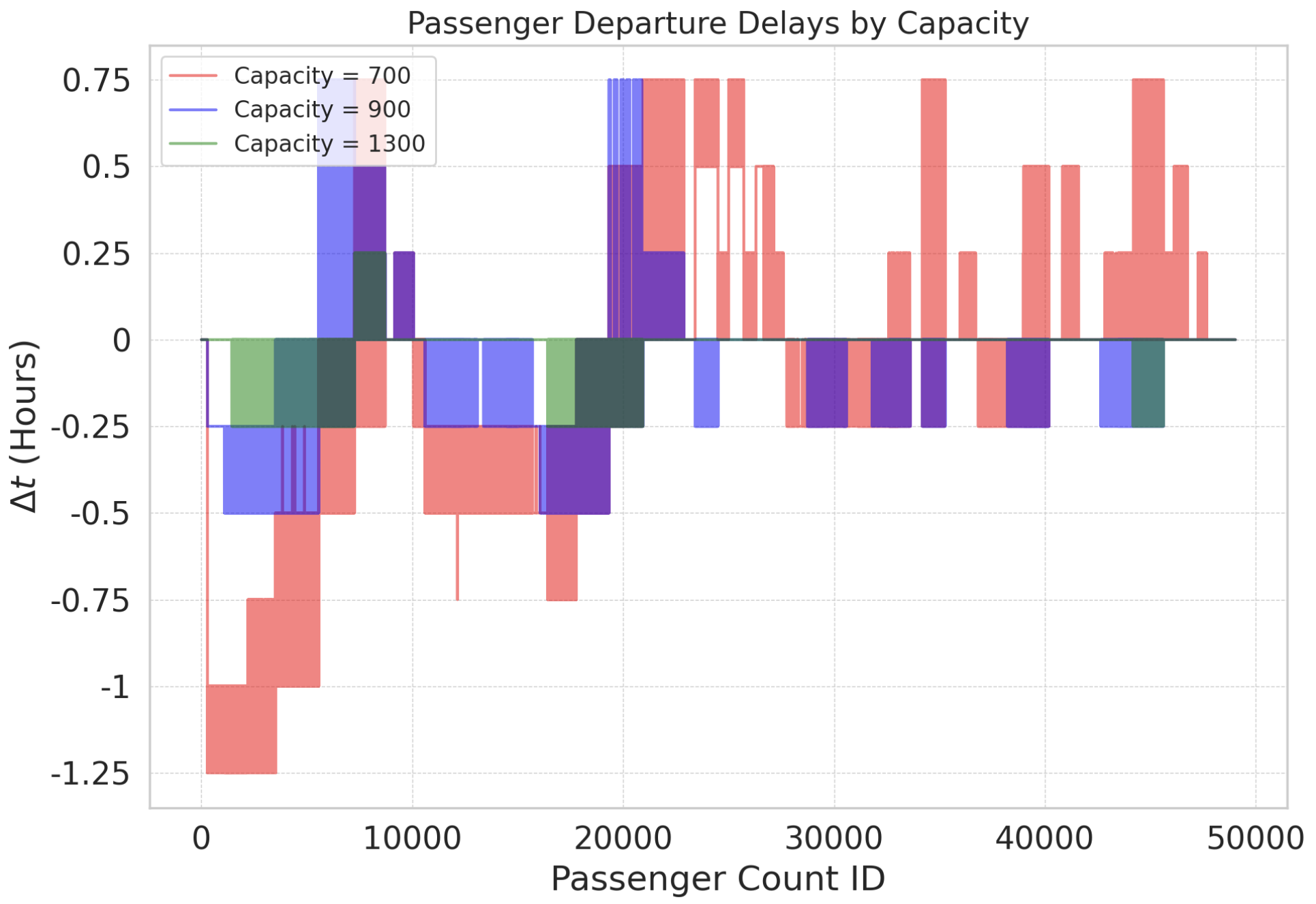}
    \caption{Variation in the deviation of the arrival-times of passengers for 3 different values of the constant capacities. Note that lesser values for the capacity constrains the system and promotes inequality, whereas the optimal solution is more equitable for all passengers for a higher value of the capacity.}
    \label{fig:constant_cap}
\end{figure}

Additionally, we examine how different capacity levels of the security affects the distribution of the reassignment across all passengers. Figure \ref{fig:constant_cap} shows the change in arrival time of passengers under this optimized system. We observe that as the capacity of the system increases, the optimization problem prefers being more equitable to all passengers as far as the reassigned arrival time is concerned. On the other hand, when the system has a lower fixed capacity, the algorithm has no choice but to ask some passengers to arrive way later in the day, whereas a few passengers are advised to arrive way too early - resulting in a more inequitable distribution of recommended arrival times. 

\begin{figure}[h]
    \centering
    \includegraphics[width=8cm, height=8cm, keepaspectratio, clip]{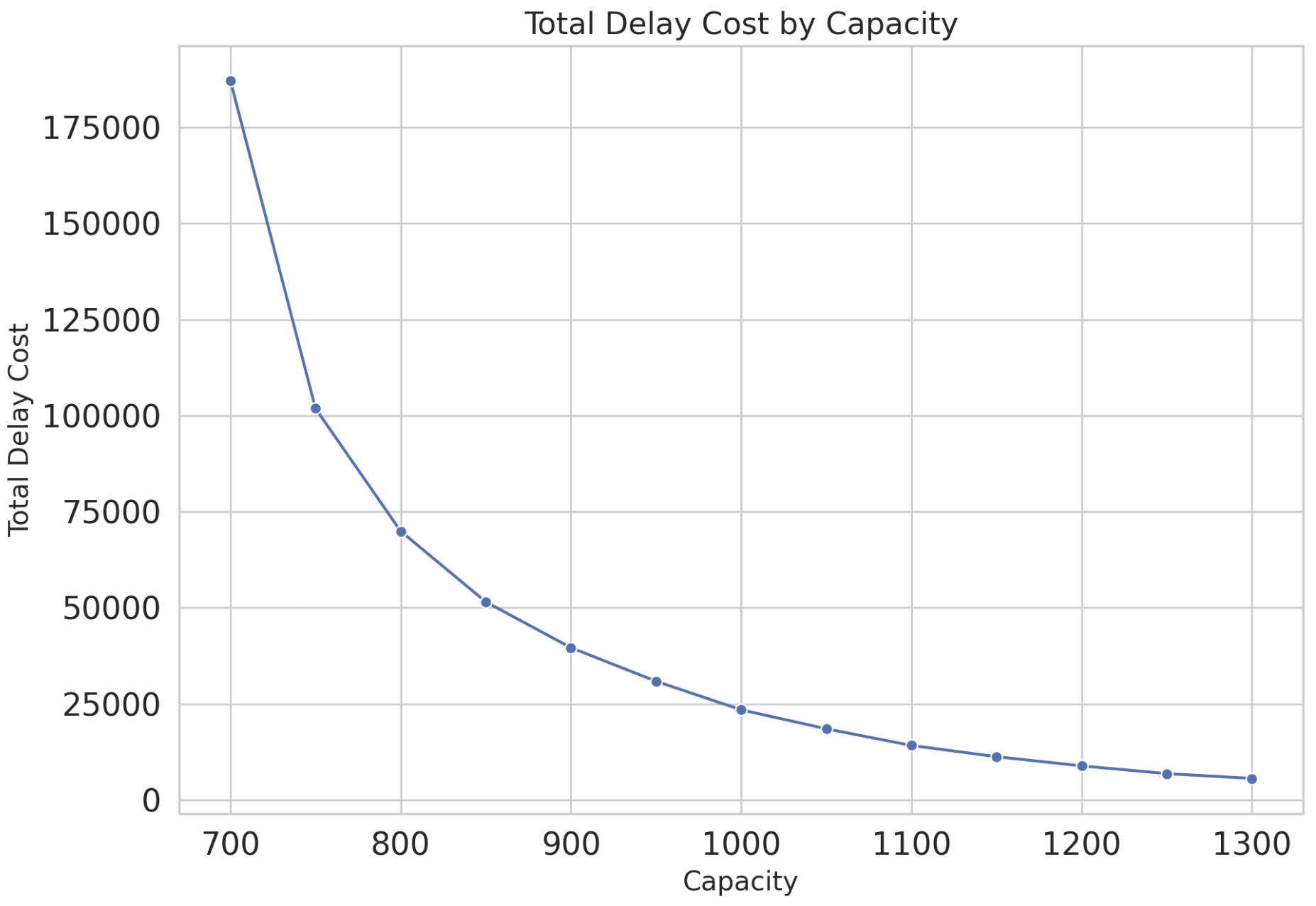}
    \caption{Increase in the capacity yields a monotonous decrease in the objective value, mainly because of the structure of the cost function.}
    \label{fig:cost_cap}
\end{figure}

Furthermore, Fig. \ref{fig:cost_cap} shows the monotonous decrease in the objective value as the capacity is increased. This is because the objective function essentially contains the cost associated with assigning a given passenger to each possible time-slot, summed up over all passengers. When the capacity of the system is higher, the cost associated with assigning a given passenger to the same time-slot decreases substantially (because of how the cost function is defined in the first place), and hence we observe a decrease in the objective value with increasing capacity.

\subsection{Time-Varying Capacity}
\label{sec:tvc}

Previously, we considered the capacity of the security system to be constant over time, and our algorithm presented the optimal reassignment for the arrival times of the passengers. Here, we propose a methodology to optimize the supply side of the system, which is the capacity at different times of the day, in order to best match the temporal demand observed at the airport security queue.

We revise the optimization problem to allow time-varying capacities at security:

\begin{align}
    \text{min} \sum_{i}\sum_{j} c_{ij}x_{ij} + \lambda_{1} \sum_{j}y_{j} + \lambda_{2} \sum_{j}^{T-1} \tau_{j}
    \label{new_of}
\end{align}
\vspace{0.3in}
\begin{align}
    \sum_{k\in V \setminus \{v\}} x_{ik} - \sum_{k\in V \setminus \{v\}} x_{ki} = b_{i} \quad \forall v \in V
\end{align}
\vspace{0.3in}
\begin{align}
    l_{ij} \leq x_{ij} \leq u_{ij} \quad \forall (i,j) \in E
\end{align}
\vspace{0.1in}
\begin{align}
    \tau_{j} \geq C_{j+1} - C_{j} \quad \forall j\in T
\end{align}
\vspace{0.1in}
\begin{align}
    \tau_{j} \geq C_{j} - C_{j+1} \quad \forall j\in T
\end{align}

\vspace{0.1in}

We retain the definition of parameters in Eq. (\ref{bi}), Eq. (\ref{lij}), and Eq. (\ref{uij}). Instead of having the security capacity as an input to the model, the time-varying capacity are now decision variables as represented by $C_{j}$. The objective function, Eq. (\ref{new_of}), is a linear combination of three terms. The first term is the reassignment cost that we defined in the original problem. The second term is the weighted cost of operating security at the determined capacity. The third term represents the sum of the absolute differences in capacities at consecutive time steps. 

While minimizing $\sum_{i}\sum_{j}c_{ij}x_{ij}$ is the primary objective, ensuring efficient passenger service with minimal staffing is achieved by minimizing the sum of individual capacities at different times. Furthermore, minimizing the absolute sum of differences between capacities at consecutive time steps ensures that capacity does not vary drastically over time. This smooth variation in staffing capacity is crucial for the staffing agency to manage the proposed distribution of capacity efficiently.

\begin{figure}[h]
    \centering
    \includegraphics[width=8cm, height=8cm, keepaspectratio, clip]{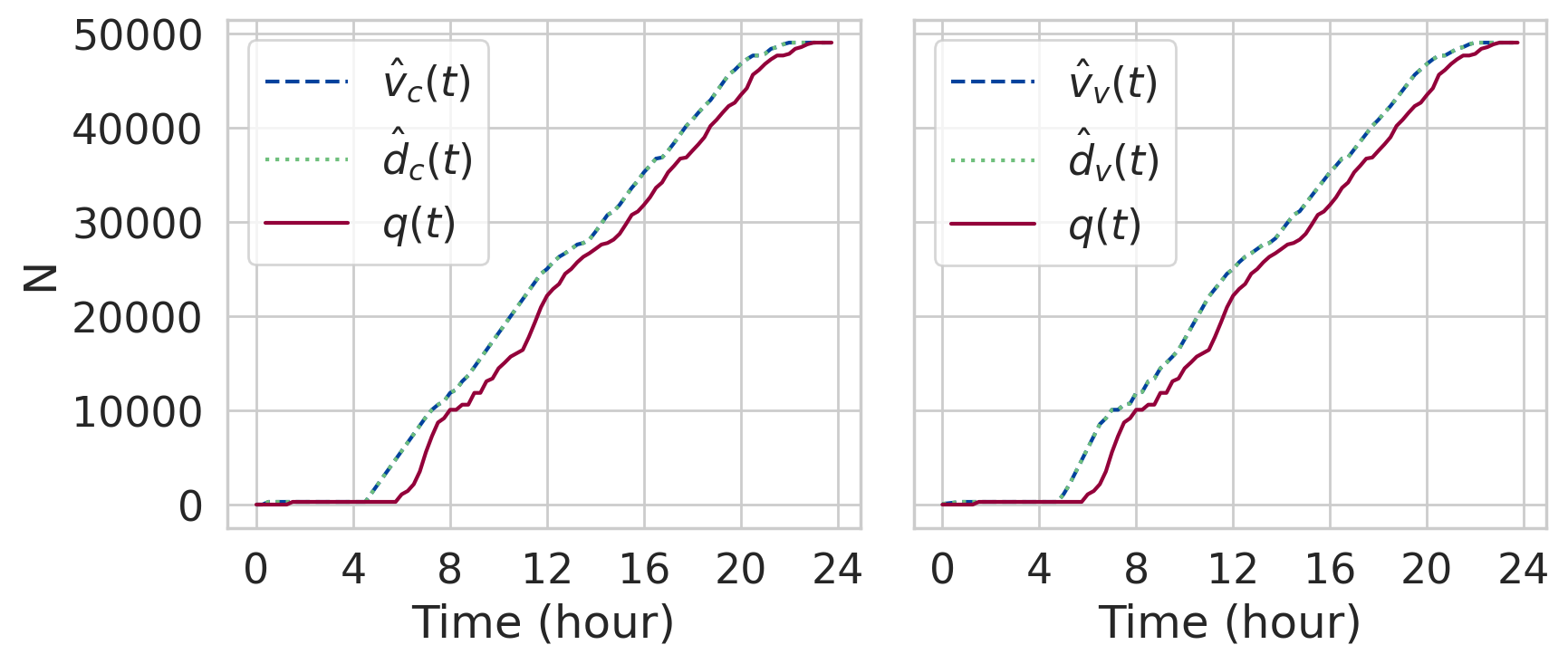}
    \caption{N-T diagram of the system under first-come first-served control (left) and under the optimal control policy (right) derived from the optimization where the capacity at different time slots can vary. The two systems yield comparable results in $TW$. In the constant capacity optimization, $C_{j} = 900$. In the time-varying capacity optimization, $\lambda_{1}=0$, and $\lambda_{2}=10$.}
    \label{fig:var_cap}
\end{figure}

\begin{figure}[h]
    \centering
    \includegraphics[width=8cm, height=8cm, keepaspectratio, clip]{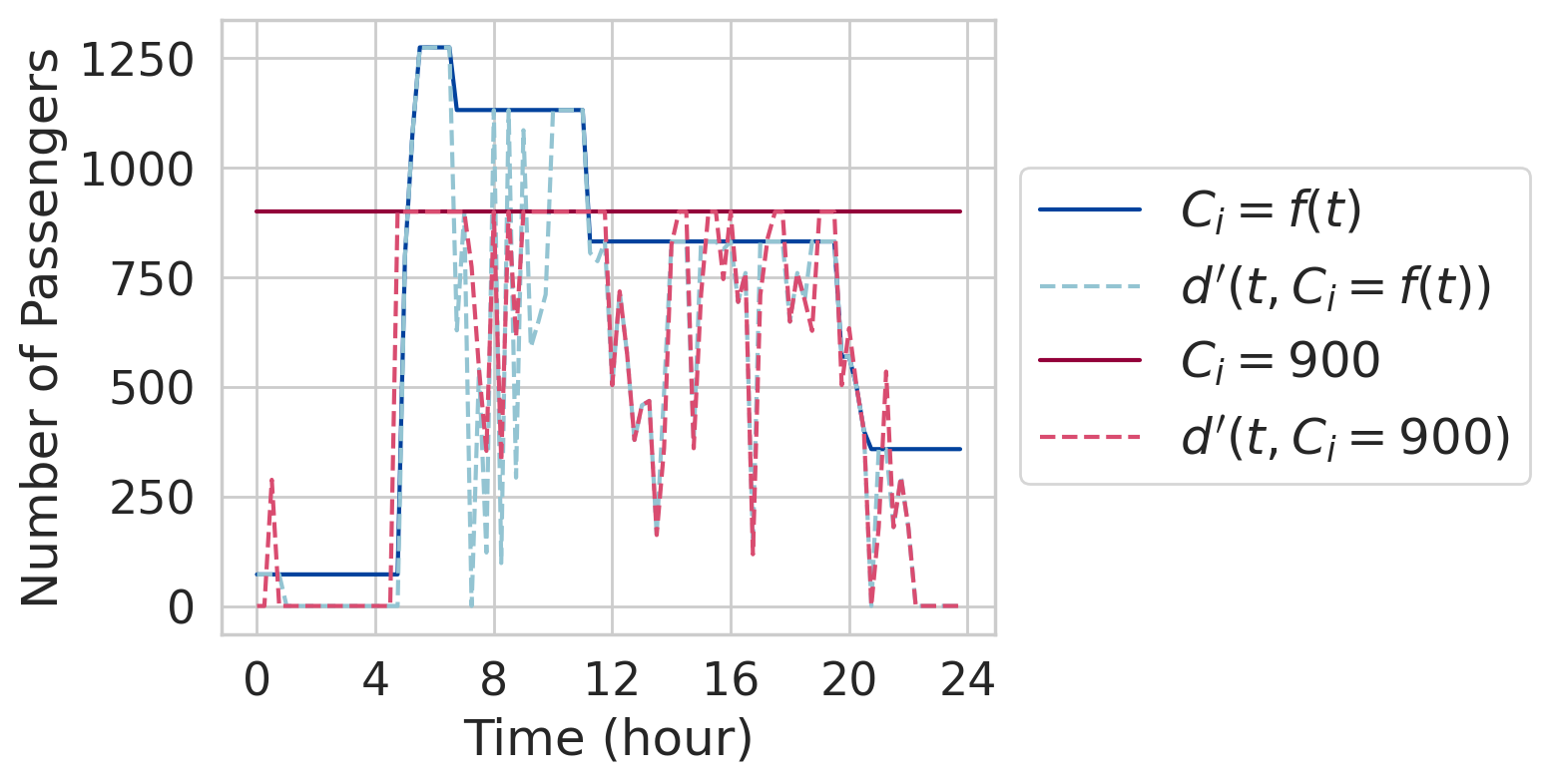}
    \caption{The capacity and the service rate in terms of the number of passengers under time-invariant and time-varying capacities. $C_{i}=f(t)$ denotes the time varying capacity determined by the optimization problem in \ref{sec:tvc}. $d^{\prime}(t, C_{i}=f(t))$ is the number of passengers completed security screening at time $t$ under the time-varying capacity. $C_{i}=900$ is the constant capacity scenario outlined in \ref{sec:const_cap} and $d^{\prime}(t, C_{i}=900)$ is the number of passengers completed security screening at time $t$ under the constant capacity. In the time-varying capacity optimization, $\lambda_{1}=0$, and $\lambda_{2}=10$.}
    \label{fig:efficiency}
\end{figure}

Figure \ref{fig:var_cap} shows the state of the first-come first-served queuing system under the constant capacity of 900 alongside the state of the system controlled with the proposed capacity as seen in Fig. \ref{fig:efficiency}. While the two systems are comparable in reducing the total waiting time of passengers in the system, optimizing for the capacity yields higher efficiency with respect to the security system. As seen in Fig. \ref{fig:efficiency}, under the constant capacity assumption, the utilization of the security system is extremely low in the early morning between 1am and 3am and after 11pm. However, with the formulation in section \ref{sec:tvc}, we lower the capacity of the security system during the low-demand period and increase the capacity when the demand is high. Note that we only have a limited number of changes in capacity, due to the regularizing term in the objective function that penalizes differences in capacity between consecutive time slots.

\subsection{Uncertainty in Compliance}
\vspace{0.1in}
Unlike the airport slot assignment problem where the decisions are implemented collectively among a limited number of parties, such as airlines and airports, the security time slot assignment is subject to great uncertainty as it is nearly impossible to forcibly align each individual passenger's behavior to reach system optimal. Therefore, one question of interest is how would the system behave if the passengers were non-compliant with the reassignment of time slots.

\begin{figure}[h]
    \centering
    \includegraphics[width=6cm, height=8cm, keepaspectratio, clip]{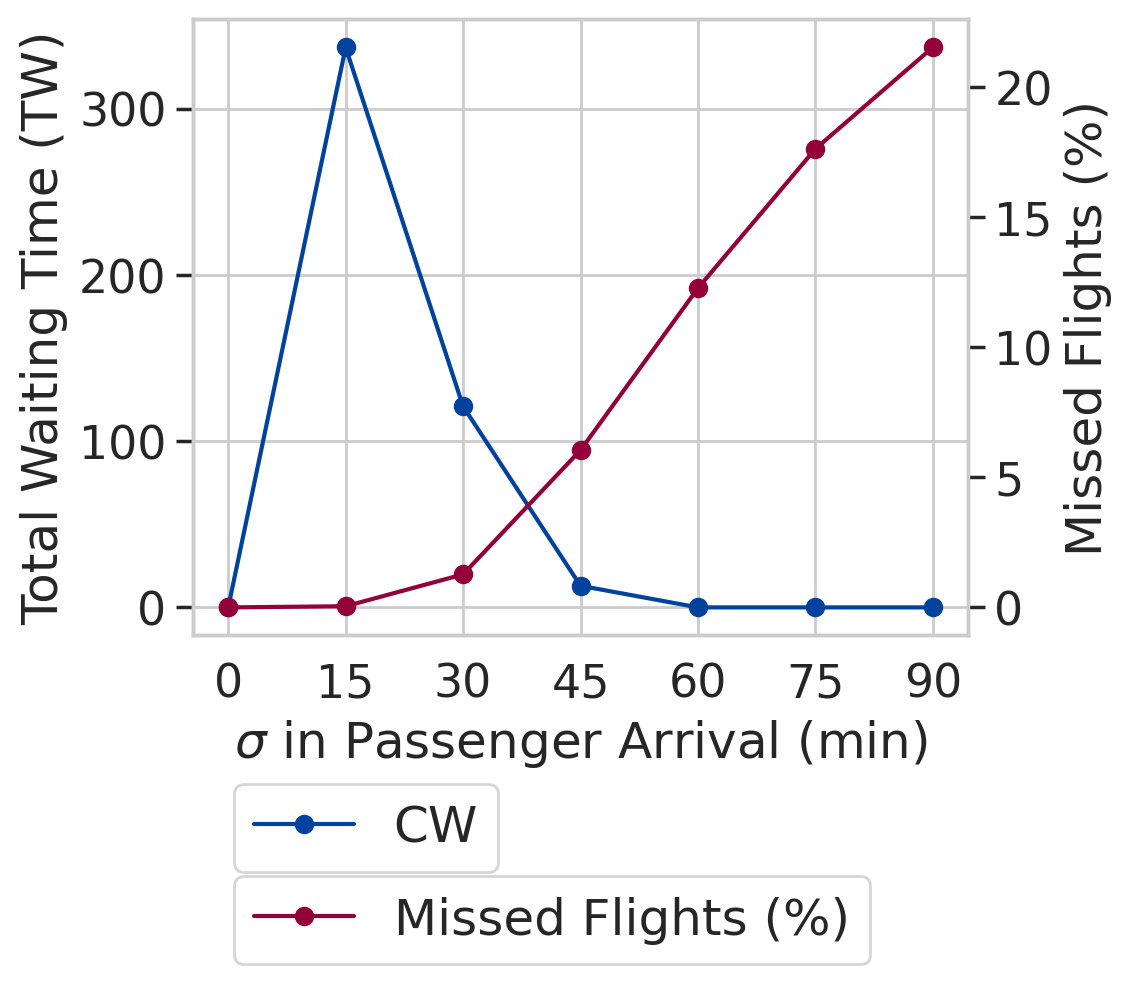}
    \caption{The relationship between total waiting time (TW), percentage of passenger missing flights, and level of deviation from assigned time slot of arrival.}
    \label{fig:comply_sigma}
\end{figure}

We identify two sources of variation in which a passenger is "non-compliant" with the reassignment. The first source of variation is by assuming all passengers accept the assigned time slot, the actual time of arrival, however, is a random process. We can model the deviation from the assigned time of arrival of these passengers as $x \sim \mathcal{N}(0, \sigma^{2})$. Let $a_{i}$ be the assigned time slot of passenger $i$. The actual time of arrival of the passenger $i$ at security is then $a_{i} + x$. By assuming a 15-minute security capacity of 900 and setting the time interval $\delta = 15$ minutes, Fig. \ref{fig:comply_sigma} shows that the variance in passenger arrival times has minimal impact on $TW$ of the system. As $\sigma$ increases, the normal distribution behaves more similar to a uniform distribution, meaning that the probability of having a certain deviation from the assignment is similar across different levels of deviation. By essentially spreading the passengers uniformly across the time horizon, we experience little to no increase in the $TW$ of the system. However, as $\sigma$ increases, passengers begin to miss flights as that their arrival time at the security is now pass the flight departure time. 

\begin{figure}[h]
    \centering
    \includegraphics[width=5.5cm, height=8cm, keepaspectratio, clip]{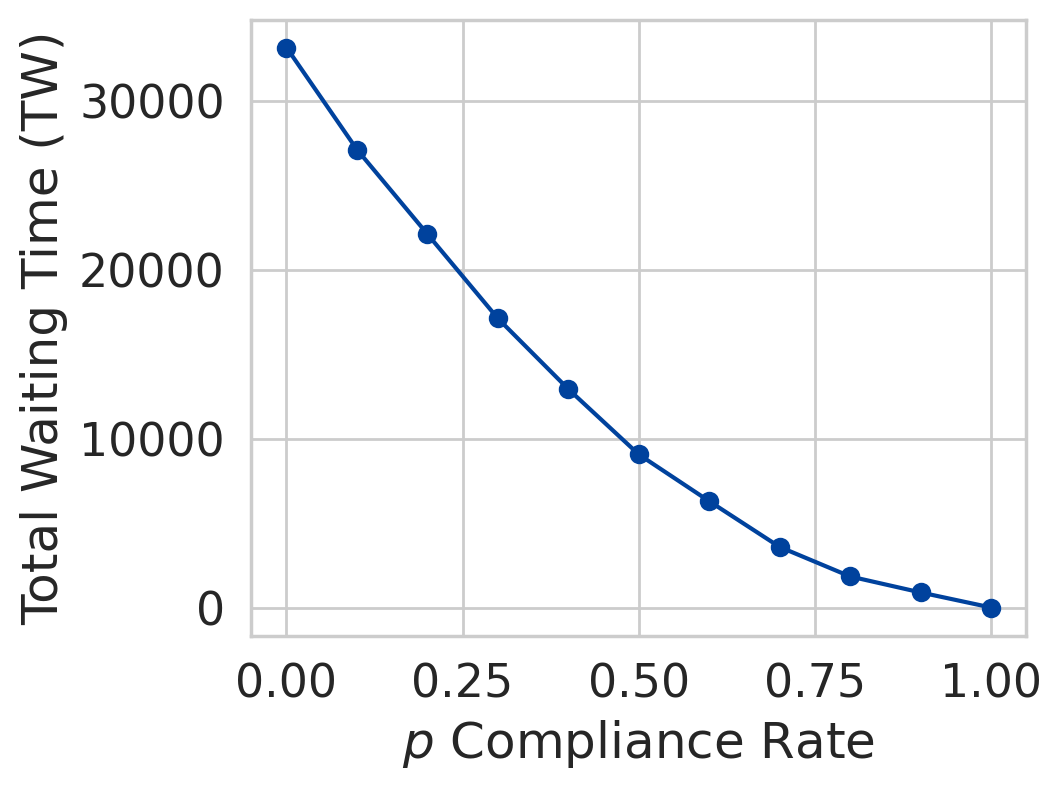}
    \caption{The relationship between total waiting time (TW) and the probability $p$ of complying with a time slot reassignment.}
    \label{fig:comply_p}
\end{figure}

The second source of non-compliance is that some passengers will reject the assigned time slot. To model this, we assume each passenger is a Bernoulli random variable with the probability of accepting a reassignment at probability $p$. If the passenger does not accept the assignment, he then arrive at the security 1 hour before the flight departure time, which is what we assumed in solving the optimization problem. Figure \ref{fig:comply_p} shows that there is an exponential decrease in $TW$ as the percentage of people accepting the assignment increases. At 50\% compliance rate, we reduce the $TW$ of the system by 63\%. The marginal gain in $TW$ reduction suggests that the slot reassignment mechanism can tolerate certain level of randomness in passenger compliance while providing improvement to the existing system.

\section{Conclusion}

In this work, we demonstrated that both the variation in actual passenger arrival times and the probability of compliance significantly impact the operational efficiency of the system. Our algorithm effectively redistributed demand based on the distribution of capacity and proposed an optimal capacity distribution to cater to the demand most efficiently. These approaches highlight the importance of considering compliance rates in optimizing system performance.

One future research direction is incorporating stochasticity to the model. The compliance rate of the passengers cannot be assumed to be 100\%. Thus, the next step in our approach is to solve a Distributed Robust Optimization (DRO) problem. The key benefit of tackling this problem using a DRO-based formulation is that the optimal solution allows for variations within a certain standard deviation, which can be fine-tuned.

We propose the following formulation for our problem as a Distributed Robust Optimization:

\begin{align}
\min_{\textbf{c}} \; & \mathbb{E}_{\xi \sim \mathcal{P}} \left[ f(\textbf{c}, \xi) \right] + \lambda \text{Var}_{\xi \sim \mathcal{P}} \left[ f(\textbf{c}, \xi) \right] \\
\text{subject to} \; & \textbf{c} \in \mathcal{C} \\
& \xi \in \mathcal{U}
\end{align}

\vspace{0.1in}

where $\textbf{c}$ represents the capacity distribution, $\xi$ denotes the uncertainty in the passenger compliance rates, $\mathcal{P}$ is the probability distribution of the uncertainty, and $\lambda$ is a regularization parameter that controls the trade-off between the expected value and the variance of the objective function. The feasible set $\mathcal{C}$ represents the constraints on the capacity, and $\mathcal{U}$ is the uncertainty set.

This approach would enhance the robustness and reliability of the system under uncertain conditions, leading to improved operational efficiency. We plan to compare the results obtained from solving this DRO problem to the solutions we already have from the MCNF based approach. This comparison will enable our objective function for the optimization problem to prioritize different terms while also allowing for a window of variance in passenger compliance.

\footnotesize

\begin{center}

\end{center}


\begin{thebibliography}{99}

\bibitem{bbc} BBC: {\sl Dublin Airport apologises after passengers miss flights due to queues}, BBC News, 2022.

\bibitem{Rosenow} Rosenow, J., Michling, P., Schultz, M., and Schönberger, J.: {\sl Evaluation of Strategies to Reduce the Cost Impacts of Flight Delays on Total Network Costs}, Aerospace, 2020, 7(11):165.

\bibitem{forbes} Stoller, G.: {\sl One of Every Seven Travelers Miss Their Flights Because of Long Airport Security Lines}, Forbes News, 2018.

\bibitem{fodors} Bhateja, A.: {\sl Expect to Lose This Much Money Every Time There’s a Major Flight Delay}, Fodors Travel Guide, 2023.

\bibitem{iata} Walsh, W.: {\sl Willie Walsh's Report on the Air Transport Industry at the 80th IATA AGM}, IATA, 2024.

\bibitem{independent} Calder, S.: {\sl Why are airport queues so long and what are passengers' rights if they miss a flight}, The Independent, 2022.

\bibitem{marshall} Marshall, Z. A., Mott, J. H., Gottwald, A. J., Patrick, C. A., Dy, L.R.: {\sl Expediting airport security queues through advanced lane assignment}, Journal of Transportation Security, 2022, Vol. 15, pp. 245-262.


\bibitem{small} Small, K. A.: {\sl Urban Transportation Economics}, Regional and Urban Economics Parts 1 \& 2, 2013, pp. 251-439.

\bibitem{vickrey} Vickrey, W. S.: {\sl Congestion Theory and Transport Investment}, The American Economic Review, 1969, Vol. 59, No. 2, pp. 251-260.

\bibitem{lamotte} Lamottea, R., Palma, A., Geroliminis, N.: {\sl On the user of reservation-based autonomous vehicles for demand management}, Transportation Research Part B, 2017, pp. 205-227.


\bibitem{odoni1} Odoni, A.: {\sl A REVIEW OF CERTAIN ASPECTS OF THE SLOT
ALLOCATION PROCESS AT LEVEL 3 AIRPORTS UNDER
REGULATION 95/93}, Massachusetts Institute of Technology, No.ICAT-2020-09, Cambridge, MA, 2020.

\bibitem{ang} Li, A., Hansen, M.: {\sl Scenario-based Strategic Flight Reassignment in Multiple Airport Regions}, International Conference on Research in Air Transportation, 2022.

\bibitem{leizhou} Zhou, L., Liang, Z., Chou, C.A., Chaovalitwongse, W.A.: {\sl Airline planning and scheduling: Models and solution methodologies}, Frontiers of Engineering Management, 2020, Vol. 7, No. 1, pp. 1-26.

\end{thebibliography}
\end{document}